\documentclass{ws-procs975x65}
\usepackage{amsmath}
\usepackage{amsfonts}
\usepackage{amssymb}
\usepackage{bm}
\usepackage{hyperref}
\usepackage{mathrsfs}
\usepackage{graphicx}

\allowdisplaybreaks
\DeclareSymbolFontAlphabet{\mathrsfs}{rsfs}
\DeclareMathAlphabet{\mathcal}{OMS}{cmsy}{m}{n}

\newcommand{\ud}{\mathrm{d}}

\newcommand{\beq}{\begin{equation}}
\newcommand{\eeq}{\end{equation}} 
\newcommand\calO{{\mathcal{O}}}

\begin{document}

\title{Analytic Approximations in GR\\and Gravitational Waves\footnote{Based on a plenary talk given at the Fifteenth Marcel Grossmann Meeting on recent developments in General Relativity, Rome, July 2018.}}
\author{Luc \textsc{Blanchet}$^{**}$}

\address{$\mathcal{G}\mathbb{R}\varepsilon{\mathbb{C}}\mathcal{O}$, Institut d'Astrophysique de Paris,\\ UMR 7095, CNRS, Sorbonne Universit{\'e}s \& UPMC Univ Paris 6,\\ 98\textsuperscript{bis} boulevard Arago, 75014 Paris, France\\
$^{**}$E-mail: luc.blanchet@iap.fr}

\begin{abstract}
Analytic approximation methods in general relativity play a very important role when analyzing the gravitational wave signals recently discovered by the LIGO \& Virgo detectors. In this contribution, we present the state-of-the-art and some recent developments in the famous post-Newtonian (PN) or slow-motion approximation, which has successfully computed the equations of motion and the early inspiral phase of compact binary systems. We discuss also some interesting interfaces between the PN and the gravitational self-force (GSF) approach based on black-hole perturbation theory, and between PN and the post-Minkowskian (PM) approximation, namely a non-linearity expansion valid for weak field and possibly fast-moving sources.
\end{abstract}


\bodymatter

%%%%%%%%%%%%%%%%% now a standard article style for the most part

\section{Methods to generate gravitational wave templates}
\label{sec:intro}

The LIGO \& Virgo detectors have opened up a fantastic new avenue in Astronomy with the discovery of gravitational waves (GWs)  generated by the orbital motion and merger of binary black hole and neutron star systems.\cite{GW150914,GW170817} This also highlights the crucial role played by analytic approximation methods in general relativity (GR), since they permit an accurate description of the two-body problem in GR, which is of direct use in the data analysis of the detectors.\cite{BuonSathya15} 

The most important method in this respect is the post-Newtonian (PN) approximation, which is an expansion when the slowness parameter $\epsilon_\text{PN}=v/c$ of the compact binary system tends to zero, where $v$ is the relative orbital velocity and $c$ the speed of light. For gravitationally bound systems such as compact binaries on quasi-circular orbits, the PN approximation comes along with the post-Minkowskian (PM) one, namely a non-linearity expansion around the Minkowski background, with small expansion parameter $\gamma_\text{PM}=G m/(r c^2)$, where $r$ is the size of the orbit and $m$ the total mass of the source. Indeed, in the bounded case we have $\gamma_\text{PM}\sim\epsilon_\text{PN}^2$. However, the most important physical application of the PM approximation is for unbound orbits, when $\gamma_\text{PN}$ and $\epsilon_\text{PM}$ are unrelated, \textit{i.e.}, the problem of scattering of ultra-relativistic particles ($\epsilon_\text{PN}\lesssim 1$) and small deviation angle. The PM approximation is sometimes called the weak-field fast-moving approximation.

Black hole perturbation theory constitutes another large body of analytic approximations in GR. In the context of compact binary systems, this approximation is important, first, for analyzing the post-merger waveform of two black holes (BHs) during the so-called ringdown phase, when the newly formed BH emits quasi-normal mode radiation, and, secondly, for describing the dynamics and GWs of asymmetric compact binaries, \textit{i.e.}, endowed with an extreme mass ratio, $\nu=m_1 m_2/(m_1+m_2)^2\ll 1$. In the latter case the perturbation method takes the more suggestive name of gravitational self-force (GSF), since it is concerned with the modifications of the background geometry of the larger BH and of the geodesic motion of the particle, due to the self field generated by the particle itself. 
\begin{figure}[h]
\begin{center}
\includegraphics[width=10cm]{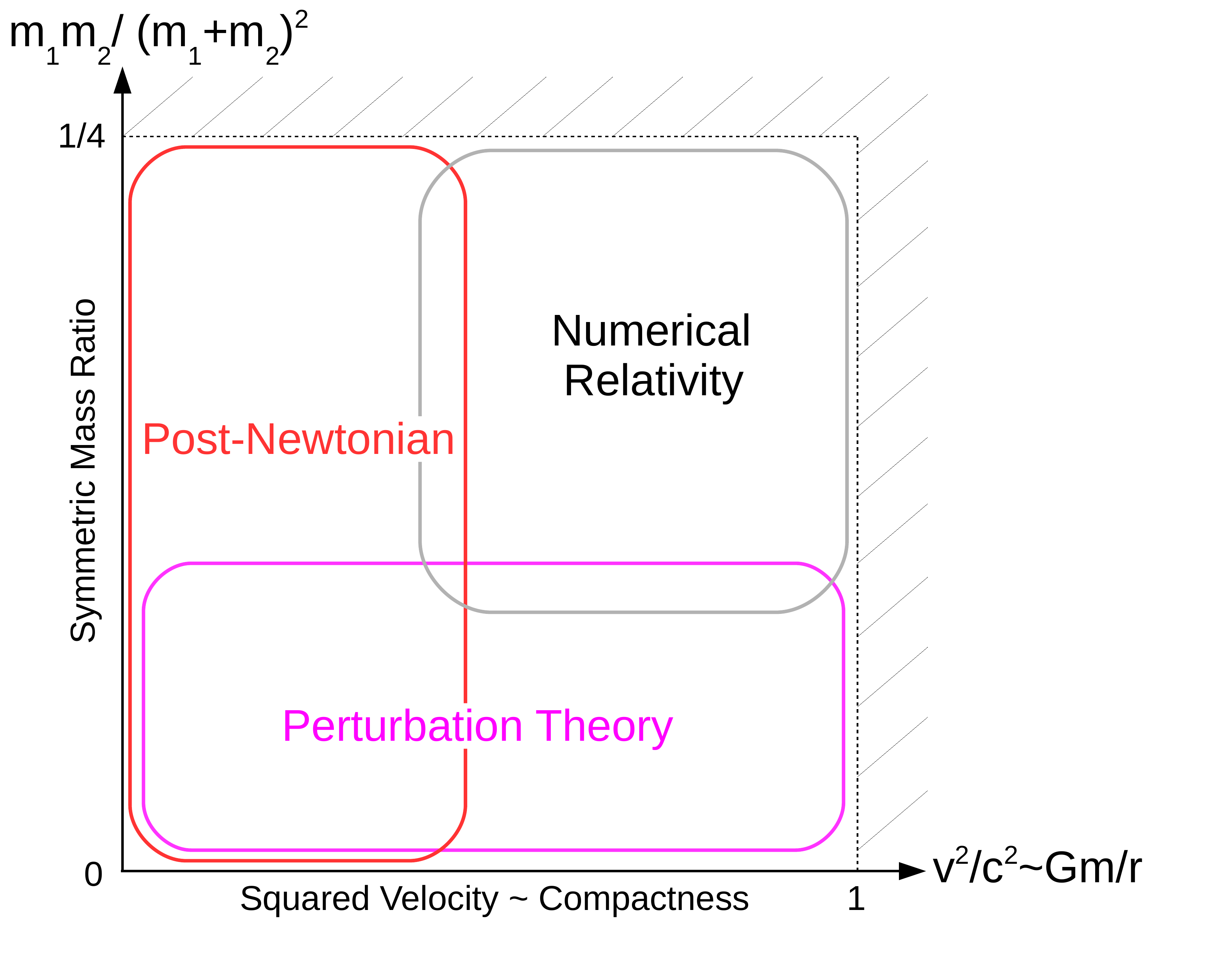}
\end{center}
\caption{Analytic approximation and numerical techniques to build GW templates for the compact binary inspiral and merger, depending on the symmetric mass ratio $\nu=m_1 m_2/m^2$ ($m=m_1+m_2$) and the slowness-weak-field parameter $\epsilon_\text{PN}=v/c\sim\sqrt{G m/r c^2}$. PN theory and perturbative GSF analysis can be compared in the slow motion weak field regime, $\epsilon_\text{PN}\ll 1$ thus $r\gg G m/c^2$, of an extreme mass ratio compact binary, $\nu\ll 1$.}\label{fig1}
\end{figure}
\begin{figure}[h]
\begin{center}
\includegraphics[width=10cm]{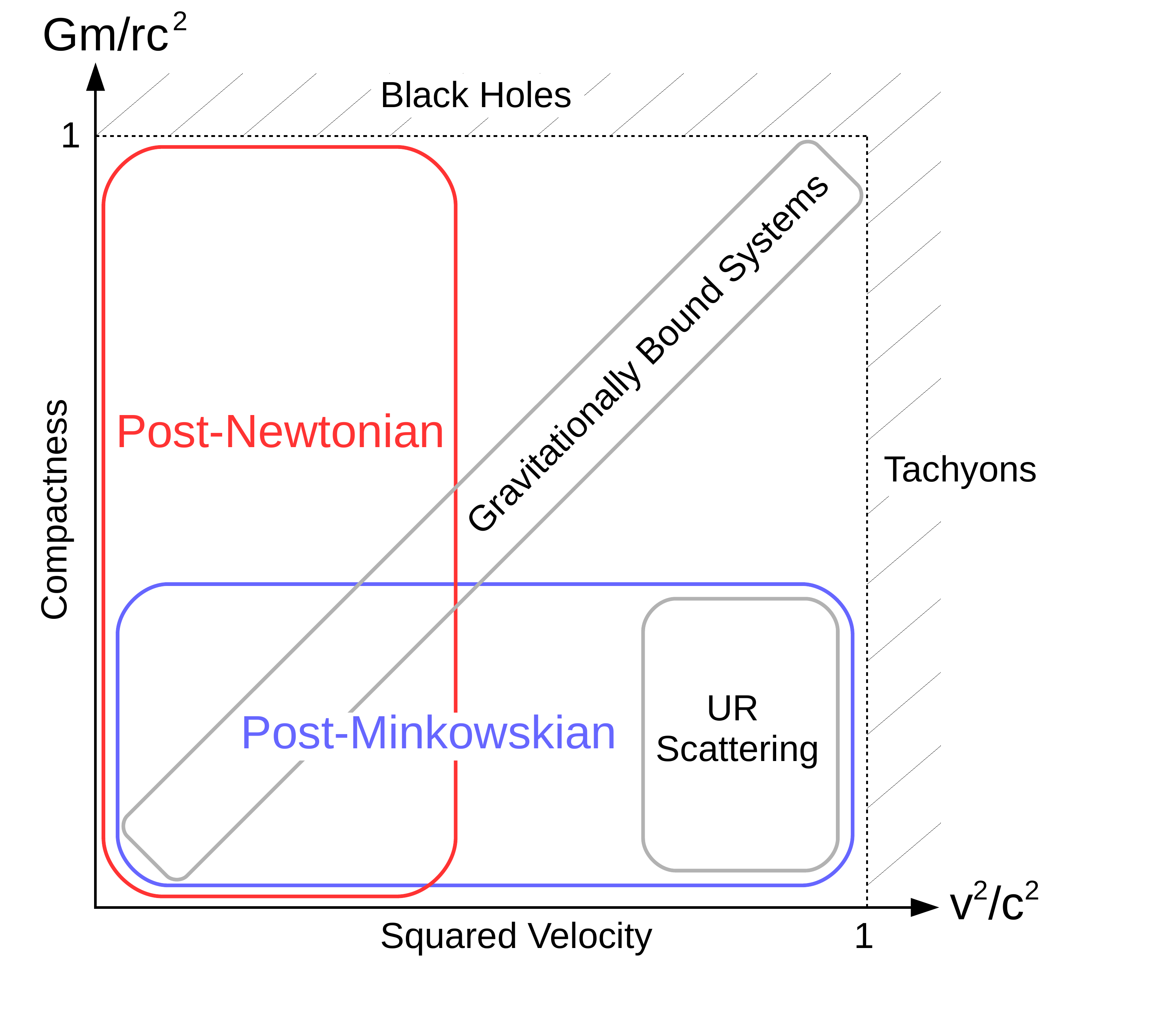}
\end{center}
\caption{Comparison between the PN and PM approximations. Gravitationally bound systems such as compact binaries on quasi-circular orbits, stand roughly on the diagonal. For such systems the PM approximation actually reduces to the PN approximation. The PM expansion is a weak-field expansion, defined with no restriction on the slowness parameter $\epsilon_\text{PN}$, and is mostly relevant in the case of unbound orbits, such as the ultra-relativistic (UR) scattering of two BHs.}\label{fig2}
\end{figure}

The domains of validity of these approximation methods, together with interesting mutual overlaps, are illustrated in Figs.~\ref{fig1} and~\ref{fig2}. Also shown in Fig.~\ref{fig1} is some comparison with numerical relativity (NR), which succeeded at solving the long standing binary black hole challenge.\cite{Pret05,Camp06,Bak06} At first sight it could seem that NR would be able to tackle and solve the complete problem of the inspiral, merger and ringdown for two compact objects. However, in order to monitor the early inspiral of two neutron stars, thousands of orbital cycles have to be computed with high precision. Then the computing times of NR become prohibitively long, and the precision of the NR simulation will never be competitive with that provided by the PN approximation. A fact that is of uttermost importance for building GW templates (and was not \textit{a priori} obvious several years ago\cite{BCT98}), is that the overlap between PN and NR exists and is quite significant. On the other hand, when the mass ratio between the two bodies is extreme, the full NR approach is unfeasible, due to the different length scales corresponding to the very different physical sizes of the compact bodies. We shall discuss in this article the important intersections between PN and GSF as shown in Fig.~\ref{fig1}, and between PN and PM, see Fig.~\ref{fig2}.

The GW templates are defined as the theoretical prediction from GR, and weighted in the Fourier domain by the detector's spectral density of noise. The templates are cross-correlated with the detector's output, and the correlation builds up when a good match occurs between a particular template and the real signal.\cite{BuonSathya15} This technique is highly sensitive to the phase evolution of the signal, which, in PN templates of compact binary coalescence, is computed from the energy balance between the decay of the binary's energy and (minus) the GW flux. 

In principle, as there is a significant overlap between the PN and NR regimes, the templates are obtained by matching together the best PN waveform for the inspiral phase (currently known to order 3.5PN) to a highly accurate numerical waveform for the merger and ringdown phases.\cite{Jena} For low mass compact binaries, such as double neutron star systems, the detectors are mostly sensitive to the inspiral phase prior to the final coalescence, and the currently known analytical PN templates are accurate enough for detection, at least when the compact bodies have moderate spins. Thus, the data analysis of neutron star binaries events such as GW170817, is essentially based on the 3.5PN templates. 

For larger masses, like BH binary events such as GW150914, the merger occurs at lower frequency, right in the middle of the detector's frequency band. Since only a few orbital cycles before the merger are seen, the match between NR and the PN is not very good and the GW templates are essentially based on the NR results. Nevertheless it is worth pointing out that the mere zeroth-order Newtonian waveform, \textit{i.e.}, based on the Einstein quadrupole formula, gives a reasonably satisfying physical interpretation of the signal even for GW150914! 

In practice, for the data analysis of large mass BH binary events, it is important to dispose of analytic rather than numerical templates, since the cross correlations must be performed with many templates associated by many trial parameters (masses and spins). In this case the templates are based on effective analytical methods that interpolate somehow between the initial PN and final NR phases. Two such techniques play a key role in the data analysis of the BH events. One is called the hybrid inspiral-merger-ringdown (IMR or IMR-Phenom) waveform and is constructed by matching together the PN and NR waveforms in an overlapping time interval described phenomenologically.\cite{Ajith11} The other technique consists of recasting the real two-body dynamics, as given by PN theory, into a simpler one-body dynamics described in a non perturbative analytic way. The so-called effective-one-body (EOB) waveform obtained in that way extends the domain of validity of the PN approximation (because it is non perturbative) and can therefore be compared and matched to the NR waveform.\cite{BuonD99} The IMR and EOB waveforms are extensively used in the LIGO \& Virgo data analysis of the recent binary BH events.

\section{State-of-the-art on equations of motion}
\label{sec:EOM}

The equations of motion (EOM) of a self-gravitating $N$-body system are written in PN like form, up to 4PN order, using one's favorite coordinate system in GR, as
\begin{align}\label{acca}
\frac{\ud \bm{v}_a}{\ud t} &= \bm{A}^\text{N}_a + \frac{1}{c^2} \bm{A}^\text{1PN}_a + \frac{1}{c^4} \bm{A}^\text{2PN}_a + \frac{1}{c^5} \bm{A}^\text{2.5PN}_a + \frac{1}{c^6} \bm{A}^\text{3PN}_a + \frac{1}{c^7} \bm{A}^\text{3.5PN}_a + \frac{1}{c^8} \bm{A}^\text{4PN}_a \nonumber\\&+ \calO \left(
  \frac{1}{c^9} \right)\,. \end{align}
The first term is of course, the usual Newtonian acceleration of $N$ ``planets'',
\begin{equation}\label{N}
\bm{A}^\text{N}_a = - \sum_{b\neq a}\frac{G m_b}{r_{ab}^2}\bm{n}_{ab}\,.
\end{equation}
The historical works in GR started in the early days of this theory, and solved the problem of the EOM at the 1PN level beyond the Newtonian term.\cite{LD17,EIH} This famous Lorentz-Droste-Einstein-Infeld-Hoffmann 1PN correction is fully given as
\begin{align}\label{1PN} 
\bm{A}^\text{1PN}_a =& - \sum_{b\neq a}\frac{G m_b}{r_{ab}^2}\bm{n}_{ab}\biggl[ v_a^2
  +2v_b^2-4(v_{a} v_{b})-\frac{3}{2}(n_{ab}v_b)^2 \nonumber\\ &\qquad\qquad - 4 \sum_{c\neq a}\frac{G m_c}{r_{ac}} - \sum_{c\neq b}\frac{G m_c}{r_{bc}}\left(1- \frac{r_{ab}}{2r_{bc}}(n_{ab}n_{bc})\right) \biggr]\nonumber\\&
  + \sum_{b\neq a}\frac{G
    m_b}{r_{ab}^2}\bm{v}_{ab}\bigl[4(n_{ab}v_a)-3(n_{ab}v_b)\bigr] - \frac{7}{2}\sum_{b\neq a}\sum_{c\neq b}\frac{G^2
    m_b m_c}{r_{ab} r_{bc}^2}\bm{n}_{bc}\,,
\end{align}
where we denote $\bm{v}_a=\ud\bm{y}_a/\ud t$, $\bm{v}_{ab}=\bm{v}_a-\bm{v}_b$, $r_{ab}=\vert\bm{y}_a-\bm{y}_b\vert$, $\bm{n}_{ab}=(\bm{y}_a-\bm{y}_b)/r_{ab}$ and the parenthesis indicate the usual Euclidean scalar product, \textit{e.g.} $(n_{ab}v_b)=\bm{n}_{ab}\cdot\bm{v}_{b}$. 

Up to the 2PN level the system is \textit{conservative}, \textit{i.e.}, admits the ten invariants associated with the symmetries of the Poincar\'e group. The first \textit{dissipative} effect appears at the 2.5PN order and features the radiation reaction damping of the system by GW emission. The 2PN and 2.5PN approximations were motivated by the Hulse-Taylor binary pulsar and worked out at the time of its discovery.\cite{OO74a,DD81b,Dhouches,S85,BFP98,IFA01} Later the motivation for the 3PN EOM came from the development of the LIGO \& Virgo detectors and the need of accurate GW templates for inspiralling compact binaries. The 3PN EOM took some time to be fully understood and completed,\cite{JaraS98,BFeom,ABF01,DJSdim,BI03CM,BDE04,itoh1,itoh2,FS3PN} together with the relatively easier dissipative 3.5PN term.\cite{PW02,KFS03,NB05,itoh3}

Three techniques have been undertaken to obtain the 4PN EOM. One is based on the Arnowitt-Deser-Misner (ADM) Hamiltonian formalism of GR in ADM coordinates,\cite{JaraS12,JaraS13,DJS14,DJS16} and has led to complete results but for the appearance of one ``\textit{ambiguity}'' parameter. The second technique is based on the Fokker action of GR in harmonic coordinates,\cite{BBBFMa,BBBFMb,BBBFMc,MBBF17,BBFM17} and has obtained complete results, \textit{i.e.}, free of any ambiguity parameter. The third one is the effective field theory (EFT),\cite{FS4PN,FStail,GLPR16,FMSS16} which yielded partial results up to now (the terms $\propto G^4$ still being in progress), but is expected to also be free of any ambiguity parameter.\cite{PR17}

In this section we describe the approach based on the Fokker action in harmonic coordinates. We start with the gravitation-plus-matter action of GR, in which the gravitational piece includes the usual harmonic gauge-fixing term, and the matter term is that of $N$ particles without spins, and with negligible internal structure:
\begin{equation}\label{S}
S = \frac{c^{3}}{16\pi G} \int \ud^{4}x \,
  \sqrt{-g} \Bigl[ R -\frac{1}{2}
        g_{\mu\nu}\Gamma^{\mu}\Gamma^{\nu} \Bigr] - \sum_{a=1}^N m_a c^{2} \int \ud t\,\sqrt{-(g_{\mu\nu})_a
      \,v_a^{\mu}v_a^{\nu}/c^{2}}\,.
\end{equation}
Here $\Gamma^\mu=g^{\rho\sigma}\Gamma^\mu_{\rho\sigma}$ and we use, for practical calculations, the Landau-Lifshitz form of the action (\textit{i.e.}, modulo a total divergence). 

The Fokker action is obtained when we insert back into~\eqref{S} an explicit PN solution of the corresponding gauge-fixed Einstein field equations, and given by an explicit functional of the particle's trajectories, \textit{i.e.}, 
\begin{equation}\label{gbar}
\bar{g}_{\mu\nu}(\mathbf{x},t)=g_{\mu\nu}[\mathbf{x}; \bm{y}_a(t), \bm{v}_a(t), \cdots]\,.
\end{equation}
The ellipsis indicate that the metric also depends on accelerations $\bm{a}_a=\ud \bm{v}_a/\ud t$, derivatives of accelerations $\bm{b}_a=\ud \bm{a}_a/\ud t$, \textit{etc.}, since we do not perform any replacements of accelerations when iterating the Einstein field equations, the EOM being considered off-shell at this stage. Substituting $\bar{g}_{\mu\nu}$ into Eq.~\eqref{S} defines the Fokker action $S_\text{F}[\bm{y}_a, \bm{v}_a,\cdots]$, and the EOM of the self-gravitating system of particles are obtained as the (generalized Euler-Lagrange) equations
\begin{align}\label{EOM}
\frac{\delta S_\text{F}}{\delta \bm{y}_a} = 0\,.
\end{align}
Once they have been constructed, the EOM can be order reduced by replacing all the higher-order accelerations by their expressions coming from the lowest-order PN equations. The Fokker action describes only the conservative dynamics, and the dissipative effects have to be added separately in the EOM. Note that the Fokker action is equivalent to the EFT action in the ``tree-level'' approximation, in which we neglect quantum loops.
 
An interesting feature of the local (near zone) 4PN dynamics is that there is an imprint of GW tails propagating at infinity. The tails are secondary non-linear waves caused by backscattering of linear waves onto the space-time curvature generated by the total mass $M$ of the source. Part of the effect can be seen as a tail-induced modification of the leading 2.5PN radiation reaction force at the relative 1.5PN order.\cite{BD88,B93,B97} However, associated with this dissipative piece, there exists also a conservative effect which thus enters into the 4PN conservative dynamics. Its contribution to the Fokker action reads as\cite{FStail,DJS14,BBBFMa,GLPR16}
\begin{equation}\label{Stail0}
S_\text{F}^\text{tail} = \frac{G^2M}{5c^8} \int_{-\infty}^{+\infty}
\ud t \,I_{ij}^{(3)}(t) \int_0^{+\infty} \ud\tau
\ln\left(\frac{\tau}{\tau_0}\right)\left[I_{ij}^{(4)}(t-\tau) -
  I_{ij}^{(4)}(t+\tau)\right] \,.
\end{equation}
Here $I_{ij}=\sum_a m_a y_a^{\langle i} y_a^{j\rangle}$ is the Newtonian quadrupole moment of the system (the angular brackets refer to the symmetric-trace-free projection), the superscript $(n)$ denotes multiple time derivatives, and $\tau_0$ is an arbitrary constant. To leading order the total mass $M$ reduces to $\sum_a m_a$, but at higher order it should involve the contribution of the gravitational binding energy of the particles. An elegant rewriting of Eq.~\eqref{Stail0} is with the Hadamard ``\textit{Partie finie}'' (Pf) integral,
\begin{equation}\label{Stail}
S_\text{F}^\text{tail} = \frac{G^2M}{5 c^8} \,\mathop{\text{Pf}}_{\tau_0}
\int\!\!\int \frac{\ud t\,\ud t'}{\vert t-t'\vert}
I_{ij}^{(3)}(t)\,I_{ij}^{(3)}(t')\,.
\end{equation}
Due to this conservative tail contribution, the 4PN dynamics is non-local in time, and this entails subtleties in the derivation of the invariants of motion, which have been recently fully elucidated.\cite{DJS14,DJS16,BBBFMb,BL17}

The calculation crucially relies on the systematic use of dimensional regularization (DR), to cure both ultra-violet (UV) divergences due to the model of point particles adopted to describe the compact objects, and infra-red (IR) divergences that start appearing precisely at the 4PN order and are associated with GW tails. We are here borrowing DR from EFT and quantum field theory, and we use it in the classical $N$-body problem as a mean to preserve the diffeomorphism invariance of GR. For this reason we conjecture that DR is the only known regularization technique able to successfully solve the problem of EOM in high PN approximations.

In an initial calculation (valid for two particles, $N=2$), we used DR for UV divergences but a variant of the Hadamard regularization (HR) for IR divergences.\cite{BBBFMa,BBBFMb} Based on some trial calculations, using various types of regularizations, we conjectured that the results of different IR regularizations will physically differ by at most two parameters called ambiguities and denoted $\delta_1$ and $\delta_2$.\cite{BBBFMb} This finding was in agreement with an earlier suggestion.\cite{DJS16} Modulo unphysical shifts of the trajectories, the two offending ambiguity terms in the Fokker Lagrangian ($S_\text{F}=\int\ud t L_\text{F}$) turn out to appear at the difficult $G^4$ level and be of the form
\begin{equation}\label{delta12}
\delta L_\text{F} = \frac{G^4 (m_1+m_2)
    \,m_1^2m_2^2}{c^8r_{12}^4}\Bigl(\delta_1 (n_{12}v_{12})^2 +
  \delta_2 v_{12}^2 \Bigr)\,.
\end{equation}
To determine what the values of these ambiguities are we embarked on the DR treatment of the IR divergences. Consider a typical term in the Fokker Lagrangian with non-compact support and generic function $F$, which diverges at infinity. With HR such term is treated as
\begin{equation}\label{HR}
L_\text{F}^\text{HR} = \mathop{\text{{\rm FP}}}_{B=0}
  \int_{r>\mathcal{R}} \ud^{3}\mathbf{x}
  \,\Bigl(\frac{r}{r_0}\Bigr)^B F(\mathbf{x})\,.
    \end{equation}
Since we focus on IR divergences we consider only the far zone contribution $r>\mathcal{R}$, where $\mathcal{R}$ denotes an arbitrary large radius, typically the inner radius of the wave zone. The Finite Part (FP) operation is closely related to the Hadamard partie finie Pf, and consists of applying analytical continuation in 
$B\in\mathbb{C}$, expanding the integral when $B$ tends to zero, and keeping only the coefficient of the zero-th power of $B$ in that expansion (discarding any strictly positive or negative power of $B$). On the other hand, with DR the same term is treated as
\begin{equation}\label{DR}
L_\text{F}^\text{DR} = \int_{r>\mathcal{R}}
  \frac{\ud^d\mathbf{x}}{\ell_0^{d-3}}\,F^{(d)}(\mathbf{x})\,,
    \end{equation}
where $F^{(d)}$ is the $d$-dimensional analogue of the generic function $F$ in~\eqref{HR}, and where $\ell_0$ is the characteristic length scale associated with DR. We find that the difference between the two prescriptions is given by\cite{BBBFMc}
\begin{equation}\label{diff}
L_\text{F}^\text{DR} - L_\text{F}^\text{HR} = 
    \sum_q\biggl[ \frac{1}{(q-1)\varepsilon} - \ln\left(\frac{r_0}{\ell_0}\right)\biggr]
    \int\ud\Omega_{2+\varepsilon}\,\varphi^{(\varepsilon)}_{3,q}(\mathbf{n})
    + \mathcal{O}\left(\varepsilon\right)\,.
    \end{equation}
The functions $\varphi^{(\varepsilon)}_{p,q}$ represent the coefficients of $r^{-p-q\varepsilon}$ (with $p, q\in\mathbb{Z}$) in the expansion of $F^{(d)}$ when $r\to+\infty$ along the direction $\mathbf{n}$. We pose $\varepsilon = d-3$ and neglect the terms dying with $\varepsilon\to 0$. The angular integration in~\eqref{diff} is over the sphere in $d-1=2+\varepsilon$ dimensions. Notice that the result~\eqref{diff} depends only on the singular coefficients $\varphi^{(\varepsilon)}_{3,q}$ of the expansion of $F^{(d)}$ at infinity, and that the arbitrary scale $\mathcal{R}$ has disappeared from it. 

The formula~\eqref{diff} contains an IR pole $\propto 1/\varepsilon$. In the language of the EFT, this pole comes from the ``potential mode'' contribution. However, we have also to take into account the 4PN tail effect given by Eq.~\eqref{Stail} but in $d$ dimensions. In the EFT language this will correspond to the ``radiation'' contribution and should also contain a pole $\propto 1/\varepsilon$, but this time of UV type. Our explicit calculation has shown that the two IR and UV poles exactly cancel out (modulo unphysical shifts of the trajectories),\cite{BBBFMc,MBBF17} in complete agreement with general arguments within the EFT.\cite{PR17} The 4PN tail term in $d$ dimensions takes the same form as~\eqref{Stail} but with the arbitrary scale $\tau_0$ determined by the DR scale $\ell_0$ as
\begin{equation}\label{tau0}
\tau_0^\text{DR} = \frac{2\ell_0}{c
  \sqrt{4\pi}}\,\text{exp}\Bigl[\frac{1}{2\varepsilon} - \frac{1}{2}\gamma_\text{E} - \frac{41}{60}\Bigr]\,,
  \end{equation}
with $\gamma_\text{E}$ denoting the Euler constant, and the UV pole $\propto 1/\varepsilon$ cancelling the IR one in~\eqref{diff}. The result~\eqref{tau0} has been obtained thanks to a ``matching'' equation relating the near zone which is the domain of validity of the PN approximation, to the far zone where GW tails propagate. Finally we find that the rational fraction $-\frac{41}{60}$ in~\eqref{tau0}, is just the one necessary and sufficient to determine the values of the two ambiguity parameters in~\eqref{delta12} as
\begin{equation}\label{delta12res}
\delta_1 = -\frac{2179}{315}\,,\qquad\delta_2 = \frac{192}{35}\,,
\end{equation}
therefore resolving the problem of ambiguities. The values~\eqref{delta12res} are consistent with numerical and analytical GSF calculations of the energy and periastron advance for circular orbits in the small mass ratio limit.\cite{DJS14,DJS16,BBBFMc} Remarkably, the result~\eqref{tau0} agrees with that of Galley \textit{et al.},\cite{GLPR16} obtained by means of a diagrammatic evaluation of the tail term in $d$ dimensions with EFT methods. On the other hand, the lack of a consistent matching between the near and far zones in the ADM Hamiltonian formalism,\cite{JaraS12,JaraS13,DJS14,DJS16} and therefore a complete control of the tail term~\eqref{Stail} including the final determination of Eq.~\eqref{tau0}, forces this formalism to be still plagued by one ambiguity parameter, denoted $C$ in\cite{DJS14}.

\section{State-of-the-art on GW generation}
\label{sec:GW}

The two basic ingredients in the theoretical PN analysis correspond to the two sides of the energy balance equation obeyed by the binary's orbital frequency and phase. Since the orbit will have circularized by radiation reaction at the time when the signal enters the detectors' bandwidth there is no need to invoke the balance equation for the orbital angular momentum. Thus we just impose
\begin{equation}\label{balanceE}
\frac{\ud E}{\ud t} = - \mathcal{F} \,.
\end{equation}
The energy $E$ is nothing but the Noetherian conserved energy $E$ associated with the Fokker Lagrangian computed in the previous section. On the other hand, the GW energy flux $\mathcal{F}$ on the right-hand side is obtained from a GW generation formalism. From Eq.~\eqref{balanceE} one deduces the time evolution of the binary's orbital frequency $\omega$ and orbital phase $\phi$ by solving
\begin{equation}\label{phase}
\phi = \int \omega\,\ud t = - \int \frac{\omega}{\mathcal{F}}\,\frac{\ud E}{\ud\omega}\,\ud\omega \,.
\end{equation}
At the 4.5PN order, for circular orbits, the conserved energy function is given by
\begin{align}\label{E45PN}
E &= -\frac{m\,\nu c^2 x}{2} \biggl\{ 1 + \left( - \frac{3}{4} -
        \frac{\nu}{12} \right) x + \left( - \frac{27}{8} +
        \frac{19}{8} \nu - \frac{\nu^2}{24} \right) x^2 
        \nonumber\\ &\qquad\quad + \left( - \frac{675}{64} + \biggl[
          \frac{34445}{576} - \frac{205}{96} \pi^2 \biggr] \nu -
        \frac{155}{96} \nu^2 - \frac{35}{5184} \nu^3 \right) x^3
        \nonumber \\ &\qquad\quad + \left( - \frac{3969}{128} +
        \left[-\frac{123671}{5760}+\frac{9037}{1536}\pi^2 +
          \frac{896}{15}\gamma_\text{E}+ \frac{448}{15} \ln(16
          x)\right]\nu\right.\nonumber\\ & \qquad\qquad\quad \left.+
        \left[-\frac{498449}{3456}+\frac{3157}{576}\pi^2\right]\nu^2
        +\frac{301}{1728}\nu^3 + \frac{77}{31104}\nu^4\right) x^4 
        \biggr\} \,,
\end{align}
where $m=m_1+m_2$ is the total mass, $\nu=m_1 m_2/m^2$ is the symmetric mass ratio, $\gamma_\text{E}$ is Euler's constant, and we employ for convenience the PN ordering parameter $x=(\frac{G m \omega}{c^3})^{2/3}$ defined from the orbital frequency $\omega$ of the circular orbit, and which constitutes an invariant in a large class of coordinate systems. Notice that Eq.~\eqref{E45PN} is valid up to the 4.5PN order \textit{included}, as there is no term at the 4.5PN order in the conserved energy for circular orbits.

The most complete formula for the GW flux is valid at the 3.5PN order beyond the Einstein quadrupole formula. However, the 4.5PN coefficient is also known,\cite{MBF16} while the 4PN coefficient is in progress. This formula has been obtained by application of a GW generation formalism based on a Multipolar-Post-Minkowskian (MPM) expansion for the external field of an isolated source,\cite{BD86,B87,BD88,BD92} and followed by a matching to the inner (near zone) PN field of that source.\cite{B95,B98mult,PB02,BFN05} The first important step in this computation is the obtention of the multipole moments of the source, $I_{i_1\cdots i_\ell}$ (mass type) and $J_{i_1\cdots i_\ell}$ (current type). The most difficult of these moments (because it necessitates the highest PN precision) is the mass quadrupole moment $I_{i_1i_2}$, given at 3.5PN order for quasi-circular orbits as
\begin{equation}\label{Iij}
I_{i_1i_2} = m \nu \left(A \, x_{\langle i_1i_2 \rangle}+B \,
  \frac{r^2}{c^2}v_{\langle i_1i_2 \rangle} + \frac{G^2
    m^2\nu}{c^5r}\,C\,x_{\langle i_1}v_{i_2 \rangle}\right) \,,
\end{equation}
where the terms are explicitly given by\cite{BIJ02,BI04mult}
\begin{subequations}\label{IijABC}
  \begin{align}
A &= 1 + \gamma \left(-\frac{1}{42} - \frac{13}{14}\nu \right) + \gamma^2 \left(-\frac{461}{1512} - \frac{18395}{1512}\nu - \frac{241}{1512} \nu^2\right) \nonumber\\
& + \gamma^3 \left(\frac{395899}{13200} - \frac{428}{105} \ln \left( \frac{r}{r_0}\right) + \left[ \frac{3304319}{166320} -  \frac{44}{3} \ln \left(\frac{r}{{r'}_0} \right) \right] \nu \right.\nonumber\\&\qquad\qquad\qquad\qquad \left.+ \frac{162539}{16632} \nu^2 + \frac{2351}{33264}\nu^3\right)\,, \\
%%%%%%%%%%%%%%%%%%%%%%%%%%%%%%%%%%%%%%%%%%%%%%%%%%%%%%%%%%%%%%%%%%%%%
B &= \frac{11}{21} - \frac{11}{7} \nu + \gamma \left(
    \frac{1607}{378} - \frac{1681}{378} \nu + \frac{229}{378} \nu^2
    \right) \nonumber\\ &\quad + \gamma^2 \left( - \frac{357761}{19800} +
    \frac{428}{105} \ln \left( \frac{r}{r_0} \right) -
    \frac{92339}{5544} \nu + \frac{35759}{924} \nu^2 +
    \frac{457}{5544} \nu^3 \right)\,, \\
%%%%%%%%%%%%%%%%%%%%%%%%%%%%%%%%%%%%%%%%%%%%%%%%%%%%%%%%%%%%%%%%%%%%%%
C &= \frac{48}{7} + \gamma \left(-\frac{4096}{315} - \frac{24512}{945}\nu \right)\,.
  \end{align}
\end{subequations}
Here the PN ordering parameter is $\gamma=\frac{G m}{r c^2}$ where $r$ is the separation distance in harmonic coordinates. Note the two constant scales entering the logarithmic terms at the 3PN order, one being the length scale $r_0$ coming from the MPM algorithm,\cite{BD86} while the other one $r'_0$ comes from the 3PN EOM in harmonic coordinates.\cite{BFeom} 

The second step is the relationship between the multipole moments of the source, and the so-called ``radiative'' multipole moments parametrizing the observable GW at future null infinity. Such relationship involves in particular the well-known tail effects and their iterations. At the 4.5PN order the radiative mass quadrupole moment $U_{i_1i_2}$ is related to the mass quadrupole moment of the source $I_{i_1i_2}$ by
\begin{align}\label{Uij}
U_{i_1i_1}(t) &= I^{(2)}_{i_1i_2}(t) + \frac{GM}{c^3} \int^{+\infty}_0 \ud
\tau \,I^{(4)}_{i_1i_2} (t-\tau) \left[ 2\ln \left( \frac{c\tau}{2b_0}
  \right) + \frac{11}{6} \right] \nonumber\\& +
\frac{G^2M^2}{c^6} \int^{+\infty}_0 \ud \tau \,I^{(5)}_{i_1i_2}(t-\tau)
\left[ 2\ln^2 \left( \frac{c\tau}{2b_0} \right) + \frac{11}{3} \ln
  \left( \frac{c\tau}{2b_0} \right)
  \right.\nonumber\\&\left. \qquad\qquad\qquad\qquad - \frac{214}{105}
  \ln \left( \frac{c\tau}{2r_0} \right) + \frac{124627}{22050} \right]
\nonumber\\&+ \frac{G^3M^3}{c^9} \int_{0}^{+\infty} \ud \tau
\,I^{(6)}_{i_1i_2}(t - \tau) \left[\frac{4}{3} \ln^3
  \left(\frac{c\tau}{2 b_0} \right) + \frac{11}{3} \ln^2
  \left(\frac{c\tau}{2 b_0} \right) \right. \nonumber
  \\ &\left. \qquad\qquad\qquad\qquad + \frac{124627}{11025}
  \ln\left(\frac{c\tau}{2b_0}\right) -\frac{428}{105}
  \ln\left(\frac{c\tau}{2b_0}\right) \ln
  \left(\frac{c\tau}{2r_0}\right) \right. \nonumber
  \\ &\left. \qquad\qquad\qquad\qquad - \frac{1177}{315}
  \ln\left(\frac{c\tau}{2r_0}\right) + \frac{129268}{33075} +
  \frac{428}{315}\pi^2\right] \,.
\end{align}
For simplicity, we have not included here the non-linear memory effect which arises at 2.5PN order,\cite{Chr91,WW91,BD92,Th92} as well as many instantaneous (non-tails) terms, that are relatively easy to compute. The terms at 1.5PN, 3PN and 4.5PN orders shown in~\eqref{Uij} correspond to what can rightly be called the ``tail'', the ``tail-of-tail'', and the ``tail-of-tail-of-tail'', respectively.\cite{MBF16} The expression~\eqref{Uij} contains still another arbitrary scale $b_0$, parametrizing the coordinate transformation between harmonic coordinates and radiative coordinates. We find that the scale $b_0$ as well as the two previous scales $r_0$ and $r'_0$ in Eq.~\eqref{Iij} cleanly cancel out in the GW flux, expressed in terms of the invariant PN parameter $x$, which is finally given by\cite{BDIWW95,BDI95,BIWW96,BIJ02,BFIJ02,BI04mult,BDEI04,BDEI05dr}
\begin{align}\label{GWflux}
  \mathcal{F} &= \frac{32c^5}{5G}\nu^2 x^5 \biggl\{ 1 +
  \left(-\frac{1247}{336} - \frac{35}{12}\nu \right) x + 4\pi x^{3/2}
  \nonumber\\ & \quad \quad \quad + \left(-\frac{44711}{9072} +
  \frac{9271}{504}\nu + \frac{65}{18} \nu^2\right) x^2 +
  \left(-\frac{8191}{672}-\frac{583}{24}\nu\right)\pi x^{5/2}
  \nonumber \\ & \quad \quad \quad +
  \left(\frac{6643739519}{69854400}+
    \frac{16}{3}\pi^2-\frac{1712}{105}\gamma_\text{E} -
    \frac{856}{105} \ln (16\,x) \right.  \nonumber \\ & \quad \qquad
    \qquad \qquad + \left. \left[-\frac{134543}{7776} +
    \frac{41}{48}\pi^2 \right]\nu - \frac{94403}{3024}\nu^2 -
    \frac{775}{324}\nu^3 \right) x^3 \nonumber \\ & \quad \quad
  \quad + \left(-\frac{16285}{504} + \frac{214745}{1728}\nu +
  \frac{193385}{3024}\nu^2\right)\pi x^{7/2} + F_\text{4PN}\, x^4 
\nonumber \\ & \quad
  \quad \quad + \left( \frac{265978667519}{745113600} -
  \frac{6848}{105}\gamma_\text{E} -
  \frac{3424}{105}\ln\left(16x\right) + \left[ \frac{2062241}{22176} +
    \frac{41}{12}\pi^2\right]\nu \right.  \nonumber \\ & \quad \qquad
    \qquad \qquad \left. - \frac{133112905}{290304}\nu^2 -
  \frac{3719141}{38016}\nu^3 \right)\pi x^{9/2} \biggr\}\,. 
\end{align}
This is valid up to 4.5PN order, with the notable exception that the 4PN coefficient, denoted $F_\text{4PN}$ in~\eqref{GWflux}, is not yet known. However, from BH perturbation theory we know already the test mass limit of this coefficient, \textit{i.e.}, in the small mass ratio limit $\nu\to 0$:\cite{Sasa94,TSasa94,TTS96,Fuj14PN,Fuj22PN} 
\begin{align}\label{terme4PN}
F_\text{4PN} =& - \frac{323105549467}{3178375200} + \frac{232597}{4410}\gamma_\text{E} - \frac{1369}{126}\pi^2 \nonumber\\&+ \frac{39931}{294}\ln 2 - \frac{47385}{1568}\ln 3 + \frac{232597}{8820} \ln x + \mathcal{O}\left(\nu\right)\,.
\end{align}
Of course, this nice result from BH perturbation theory will have to be confirmed by PN theory, which will also be able to provide the mass ratio corrections $\mathcal{O}(\nu)$.

\section{PN theory versus GSF theory}
\label{GSF}

The conservative dynamics and GWs of compact binary systems in the extreme mass ratio limit, is the realm of the perturbative gravitational self force (GSF) theory.\cite{dWB60,MiSaTa,QuWa,DW03,GW08,Pound10} For the conservative dynamics, a comparison between GSF computations and traditional PN calculations was initiated some years ago,\cite{Det08} applying to systems that are at once slowy moving and with extreme mass ratio, see the overlap region between PN and perturbation theory in Fig.~\ref{fig1}. In recent years the possibility for this comparison has been dramatically extended. Such progress is due in large part due to high precision numerical and analytical computations from a self force perspective,\cite{MST96a,MST96b,MT97,SBD08,BiniD13,BiniD14a,BiniD14b,SFW14} and to extensive analytical computations within the PN approximation.\cite{BDLW10a,BDLW10b,LBW12,BFW14a,BFW14b}

For a particle moving on an exact circular orbit around a Schwarzschild BH (neglecting radiation reaction), one disposes of a very interesting quantity, which is the invariant associated with the helical Killing symmetry, appropriate for exact circular orbits.\cite{Det08} This invariant, denoted $u_1^t$, is defined by
\begin{equation}\label{utdef}
u_1^\mu = u_1^t \,K_1^\mu\,,
\end{equation}
where $u_1^\mu$ is the normalized four-velocity of the particle 1 (with mass $m_1\ll m_2$), $K^\mu$ is the helical Killing vector (HKV) and $K_1^\mu$ the HKV at the location of the particle. Adopting a coordinate system in which the HKV reads $K^\mu\partial_\mu = \partial_t + \omega\,\partial_\varphi$, where $\omega$ is the orbital frequency of the circular orbit, the invariant reduces to the time component of the four-velocity hence its name $u_1^t$, and we have
\begin{equation}\label{ut}
u_1^t = \frac{1}{z_1} = \biggl[- (g_{\mu\nu})_1 \frac{v_1^\mu v_1^\nu}{c^2} \biggr]^{-1/2} \,,
\end{equation}
where $(g_{\mu\nu})_1$ is the metric evaluated at the particle's
location, following a certain regularization (here $v_1^\mu = \ud y_1^\mu/\ud t$ denotes the coordinate velocity, \textit{i.e.}, $y_1^0=c t$ and $v_1^0=c$). The inverse of $u_1^t$ appears to be a redshift $z_1$, and sometimes $u_1^t$ itself is called the redshift. [For a generalization of the notion of redshift to eccentric orbits, see\cite{BarackS11}.] In the exact test mass limit $\nu=0$, the invariant reduces to the one appropriate to a Schwarzschild BH,
\begin{equation}\label{utschw}
u_\text{Schw}^t = \frac{1}{\sqrt{1-3y}}\,,
\end{equation}
where $y=(\frac{G m_2\omega}{c^3})^{2/3}$ is the frequency-related PN parameter associated with the larger BH mass $m_2$. The GSF part is then defined as the coefficient of the mass ratio correction beyond Eq.~\eqref{utschw}, 
\begin{equation}\label{utSchw}
u_1^t = u_\text{Schw}^t + \nu\,u_\text{GSF}^t + \mathcal{O}(\nu^2)\,.
\end{equation}
It is clear that with this approximation the symmetric mass ratio $\nu=\frac{m_1 m_2}{m^2}$ can be replaced by the ordinary mass ratio $q=\frac{m_1}{m_2}$. In the PN approximation, the GSF part of the redshift factor appears to be an infinite PN series of the type
\begin{equation}\label{utGSF}
u_\text{GSF}^t = \sum_{j = 0}^{+\infty} \bigl(\alpha_j + \beta_j \ln y\bigr)\,y^{j+1}\,.
\end{equation}
We have included terms linear in the logarithm of $y$, but we neglect (just for this discussion) the higher powers of $\ln y$, that occur at large PN orders. Recall that the most general structure of the PN expansion involves any (integer) powers of the logarithm, $\sim (\ln y)^k y^{j+1}$.\cite{BD86} 

Here we report the results that have been obtained so far using the ``traditional'' PN approach. Recall that the PN method heavily relies on dimensional regularization (DR) to treat both UV and IR divergences, see Sec.~\ref{sec:EOM}. Another feature of the PN calculation is that it requires a machinery of tails and related non-linear effects, see Eq.~\ref{Uij}. In the PN approach one computes $u_1^t$ as a redshift in harmonic coordinates using Eq.~\eqref{ut}, and evaluates the metric at the particle's location with DR. In that way the GSF redshift has been obtained up to 4PN order as\cite{BDLW10a,BDLW10b,LBW12,BiniD13}
\begin{align}\label{ut4PN}
u^t_\mathrm{GSF} =&- y - 2 y^2 - 5 y^3 + \left( - \frac{121}{3} +
\frac{41}{32} \pi^2 \right) y^4 \nonumber\\&
+ \left( - \frac{1157}{15} + \frac{677}{512}\pi^2 
- \frac{128}{5}\gamma_\text{E} -
\frac{64}{5}\ln(16 y)\right) y^5 + \mathcal{O}(y^6)\,.
\end{align}
In addition, PN theory has been able to fix the logarithmic term at the 5PN order, which is due to a subdominant tail effect, namely
\begin{equation}\label{beta5}
\beta_5 = \frac{956}{105}\,,
\end{equation}
while the coefficient $\alpha_5$ is known only from GSF methods but has not yet been checked with PN theory. The results~\eqref{ut4PN}--\eqref{beta5} are in full agreement with direct GSF computations. This constitutes a strong confirmation of the adequation of DR for traditional PN calculations, as well as of the procedure of subtraction of the singular field which is employed by GSF theory.

A feature of the PN expansion of the redshift factor at high orders is the appearance of half-integral PN approximations, say $\frac{n}{2}$PN. At first sight these terms sound surprising because the dynamics is purely conservative (exactly circular orbits with a HKV), and we are used to the fact that half-integral PN approximations like 2.5PN are associated with dissipative radiation reaction. The existence of such terms in the conservative redshift factor, starting at 5.5PN order, has been pointed out with numerical GSF methods,\cite{SFW14} and later it was proved that these terms originate from iterated non-linear tail effects, called ``tails-of-tails''.\cite{BFW14a,BFW14b} The leading 5.5PN, next-to-leading 6.5PN and next-to-next-to-leading 7.5PN coefficients in this category have been found to be
\begin{equation}\label{alpha55}
   \alpha_\frac{11}{2} = - \frac{13696}{525}\pi \,,\qquad\alpha_\frac{13}{2} =
   \frac{81077}{3675}\pi \,,\qquad\alpha_\frac{15}{2} =
   \frac{82561159}{467775}\pi \,,
\end{equation}
while the corresponding $\beta_j$'s are zero. Notice that 7.5PN is arguably the highest order ever reached by traditional PN methods. Again the PN results~\eqref{alpha55} are in full agreement with numerical and analytical results derived by GSF methods.

\section{PN versus PM}
\label{sec:PM}

The post-Minkowskian approximation has been developed in many pionneering works.\cite{BertottiP60,WG79,HG62,Port80,WH80,BeDD81} Notably the gravitational scattering angle of two relativistic particles has been obtained up to 2PM order (quadratic in $G$).\cite{Westpf85} Recently there has been a renewal of interest in the PM approximation. Ledvinka, Sch{\"a}fer and Bi\v{c}\'{a}k\cite{LSB08} obtained a closed-form expression for the Hamiltonian of $N$ particles in the 1PM approximation, and new works appeared on the gravitational scattering angle and the link between the PM expansion and the EOB formalism.\cite{Dscatt16,Dscatt17} Here we outline our own contribution,\cite{BFok18} which concerns the comparison between 1PM and the recent 4PN calculation of the EOM of compact binaries. The Fig.~\ref{fig2} showed the domain of validity of the PM approximation \textit{versus} that of the PN expansion. 

At the 1PM approximation the field equations for a system of $N$ particles in harmonic coordinates, deduced from the gauge-fixed action~\eqref{S}, read
\begin{equation}\label{boxh}
  \Box h^{\mu\nu} = \frac{16\pi G}{c^2}\sum_{a=1}^N m_a \int_{-\infty}^{+\infty}
    \ud\tau_a \,u_a^\mu u_a^\nu \delta^{(4)}(x-y_a)\,,
  \end{equation}
where $\Box$ denotes the flat space-time d'Alembertian operator, $\delta^{(4)}$ is the four-dimensional Dirac function, $y_a^\mu$ are the particle's worldlines and $u_a^\mu=\ud y_a^\mu/(c\ud\tau_a)$ their normalized four-velocities, with the special-relativistic proper time $\ud\tau_a = \sqrt{- \eta_{\mu\nu} \ud y_a^\mu \ud y_a^\nu/c^2}$. We solve Eqs.~\eqref{boxh} by means of the standard Lienard-Wiechert procedure. Adopting a parametrization by the coordinate time $t$, \textit{i.e.}, such that $y_a^\mu=(c t, \bm{y}_a)$, the retarded time $t_a^\text{ret}$ on the trajectory $a$ associated with the propagation from $a$ to the field point $x^\mu=(c t, \mathbf{x})$, is given by the implicit retardation equation $t_a^\text{ret} = t - r_a^\text{ret}/c$ with $r_a^\text{ret}=\vert\mathbf{x} - \bm{y}_a(t_a^\text{ret})\vert$. The solution of~\eqref{boxh} is then
\begin{equation}\label{lienardW}
h^{\mu\nu}(x) = - \frac{4G}{c^2} \sum_a \frac{m_a \,u_a^\mu u_a^\nu}{r_a^\text{ret} \,(k u)_a^\text{ret}}\,,
  \end{equation}
where $k_a^\mu = [x^\mu-y_a^\mu(t_a^\text{ret})]/r_a^\text{ret} = (1,\bm{n}_a^\text{ret})$ is the Minkowski null vector between $a$ and the field point, $(k u)_a^\text{ret} = k^a_\mu u_a^\mu =\gamma_a(-1+\bm{n}_a^\text{ret}\cdot\bm{v}_a/c)$ is the usual redshift factor, with $\gamma_a=u_a^0$ the Lorentz factor and $v_a^\mu=c u_a^\mu/\gamma_a=(c, \bm{v}_a)$, the velocities being computed at the retarded time $t_a^\text{ret}$. 

However, we repeatedly use the fact that the accelerations are of order $G$ and therefore their contributions in~\eqref{lienardW} will be of order $G^2$, hence negligible with the 1PM approximation. Thus, we are allowed to assume that the four velocities $u_a^\mu$ and Lorentz factors $\gamma_a$ are constant. Furthermore, neglecting terms of order $G^2$, we can solve the retardation equation 
%$t_a^\text{ret} = t - r_a^\text{ret}/c$ 
to get the retarded time $t_a^\text{ret}$, the distance $r_a^\text{ret}$, and the direction $\bm{n}_a^\text{ret}$, in terms of their current values at time $t$, \textit{i.e.}, the ``instantaneous'' distance $r_a=\vert\mathbf{x}-\bm{y}_a(t)\vert$ and direction $\bm{n}_a = [\mathbf{x}-\bm{y}_a(t)]/r_a$. In this way, Eq.~\eqref{lienardW} becomes equivalent to
\begin{equation}\label{field}
h^{\mu\nu} = - \frac{4 G}{c^2} \sum_a \frac{m_a \,u_a^\mu u_a^\nu}{r_a \sqrt{1 + (n_a u_a)^2}}\,,
\end{equation}
which is valid at any field point except at the singular locations of the particles. Nevertheless, we can easily extend its validity to the particles by using a self-field regularization. For this purpose, it is sufficient to discard the self-field contribution from the sum of particles. An explicit self-field regularization process yielding the same result was implemented in\cite{BeDD81}. 
Therefore, at the location of particle $a$, we have
\begin{equation}\label{derfielda}
(h^{\mu\nu})_a = - \frac{4G}{c^2} \sum_{b\not= a} \frac{m_b \,u_b^\mu u_b^\nu}{r_{ab}^2 \bigl[1 + (n_{ab} u_b)^2\bigr]^{1/2}}\,,
\end{equation}
where the sum runs over all particles except $a$, we pose $r_{ab}=\vert\bm{y}_a-\bm{y}_b\vert$, and denote $n_{ab}^0=0$ and $n_{ab}^i=[\bm{y}_a - \bm{y}_b]/r_{ab}$.

The EOM of the particles is just the geodesic equation, computed at the linearized order consistent with our approximation, and we obtain
\begin{align}\label{eom1}
&\frac{\ud u_a^\mu}{\ud\tau_a} = - \frac{1}{c} \sum_{b\not= a} \frac{G m_b}{r_{ab}^2 \bigl[1 + (n_{ab} u_b)^2\bigr]^{3/2}}\biggl[ (2\epsilon_{ab}^2-1)n_{ab}^\mu \\ & ~+ (2\epsilon_{ab}^2+1)\Bigl( - (n_{ab} u_a) + \epsilon_{ab} (n_{ab} u_b)\Bigr) u_a^\mu + \Bigl( 4 \epsilon_{ab} (n_{ab} u_a) - (2\epsilon_{ab}^2+1) (n_{ab} u_b) \Bigr) u_b^\mu\biggr] \,.\nonumber
\end{align}
We use $\epsilon_{ab} = - (u_a u_b)$ as a shorthand notation. Equivalently we have also the non-covariant form of the EOM (\textit{i.e.}, PN like form), in which we introduce the ordinary velocities and accelerations and the relevant Lorentz factors (with $\bm{v}_{ab}=\bm{v}_{a}-\bm{v}_{b}$),
\begin{align}\label{eom2}
\frac{\ud \bm{v}_a}{\ud t} &= - \gamma_a^{-2}\sum_{b\not= a} \frac{G m_b}{r_{ab}^2 \bigl[1 + \gamma_b^2(n_{ab} v_b)^2/c^2\bigr]^{3/2}}\biggl[ (2\epsilon_{ab}^2-1)\bm{n}_{ab} \nonumber\\ & \qquad\qquad + \gamma_b \Bigl( - 4 \epsilon_{ab} \gamma_a (n_{ab} v_a) + (2\epsilon_{ab}^2+1) \gamma_b (n_{ab} v_b) \Bigr) \frac{\bm{v}_{ab}}{c^2}\biggr]\,.
\end{align}
At the 1PM order the EOM are conservative, thus admit conserved integrals of energy, angular momentum and linear momentum. Indeed, the radiation reaction dissipative effects are at least 2PM, \textit{i.e.}, second order in $G$. The closed-form expression of the energy through 1PM reads $E = \sum_a m_a c^2 \gamma_a + V$, with the first term being the usual special-relativistic expression and
\begin{align}\label{potV}
V &= G \sum_a \sum_{b\not= a}\frac{m_a m_b}{r_{ab}\,\bigl[1 + \gamma_b^2(n_{ab} v_b)^2/c^2\bigr]^{1/2}} \Biggl\{ \gamma_a\Bigl(2\epsilon_{ab}^2+1-4\frac{\gamma_b}{\gamma_a}\epsilon_{ab}\Bigr) \\& + \frac{\gamma_b^2}{\gamma_a}\bigl(2\epsilon_{ab}^2-1\bigr) \frac{\dot{r}_{ab}(n_{ab}v_b)-(v_{ab}v_b)}{\bigl(v_{ab}^2-\dot{r}_{ab}^2\bigr)\bigl[1 + \gamma_b^2(n_{ab} v_b)^2/c^2\bigr]+\frac{\gamma_b^2}{c^2}\bigl(\dot{r}_{ab}(n_{ab}v_{b})-(v_{ab}v_b)\bigr)^2}\Biggr\}\,.\nonumber
\end{align}
We have verified\cite{BFok18} that Eqs.~\eqref{eom2} and~\eqref{potV} perfectly reproduce the PN results in harmonic coordinates, in the case of two particles ($N=2$) at the linear order in $G$ and up to the 4PN order.\cite{BBBFMc,BBFM17}

Next we consider the Lagrangian associated with the EOM~\eqref{eom1}--\eqref{eom2}, for any $N$, in harmonic coordinates. The Lagrangian will be given by the special-relativistic expression plus terms of order $G$, and again, we neglect higher-order terms in $G$. In PN theory, it is known that the Lagrangian in harmonic coordinates is a \textit{generalized} one, depending not only on positions and velocities $\bm{y}_a, \bm{v}_a$ but also on accelerations $\bm{a}_a=\ud\bm{v}_a/\ud t$.\cite{DD81b} Such accelerations are contained in terms at least linear in $G$, so that, replacing the accelerations by the EOM would yield negligible terms of order $G^2$ at least. However, it is not allowed to replace accelerations in a Lagrangian while remaining in the same coordinate system. Such replacement is equivalent to a shift in the particles' trajectories (or so-called ``contact'' transformation), \textit{i.e.}, the new Lagrangian is physically equivalent to the original one but written in a different coordinate system.\cite{S84} Furthermore, by employing the technique of double-zero (or multiple-zero) terms, it is sufficient to consider a Lagrangian that is linear in accelerations (since the procedure can work for any PN order, and is thus formally valid at the 1PM order). Therefore, we look for a Lagrangian of the form 
\begin{equation}\label{Ldef}
L\bigl[y, v, a\bigr] = - \sum_a \frac{m_a c^2}{\gamma_a} + \lambda + \sum_a q_a^i a_a^i\,.
\end{equation}
We symbolize the functional dependence by $L[y, v, a] \equiv L[\{\bm{y}_a, \bm{v}_a, \bm{a}_a\}]$. The terms $\lambda$ and $q_a^i$ are of order $G$ and depend only on positions and velocities, \textit{i.e.}, $\lambda[y, v]$ and $q_a^i[y, v]$. Denoting by $p_a^i$ and $q_a^i$ the conjugate momenta associated with the positions $y_a^i$ and velocities $v_a^i$, \textit{i.e.},
\begin{subequations}\label{paqa}\begin{align}
p_a^i =& \frac{\delta L}{\delta v_a^i} = \frac{\partial
L}{\partial v_a^i}-\frac{\ud}{\ud t} \bigg( \frac{\partial L}{\partial
a_a^i} \bigg)\,, \label{pa}\\ q_a^i =& \frac{\delta L}{\delta
a_a^i} = \frac{\partial L}{\partial a_a^i}\,,\label{qa}
\end{align}\end{subequations}
the EOM take the ordinary Euler-Lagrange form
\begin{equation}\label{eomfull}
\frac{\ud p_a^i}{\ud t} = \frac{\partial L}{\partial y_a^i}\,,
\end{equation}
while the conserved energy $E$ is given by the generalized Legendre transformation
\begin{equation}\label{conservedE}
E = \sum_a \Bigl( p_a^i v_a^i + q_a^i a_a^i \Bigr) - L\,.
\end{equation}
In both~\eqref{eomfull} and~\eqref{conservedE} we are allowed to replace the accelerations by the EOM. For instance, the term $q_a^i a_a^i$ in $E$ will be second-order in $G$ and can be neglected at 1PM order. With~\eqref{Ldef} we obtain the EOM
\begin{equation}\label{eomexpl}
f_a^i = \frac{\delta \lambda}{\delta y_a^i} + \ddot{q}_a^i \,,
\end{equation}
where $f_a^i= m_a \frac{\ud}{\ud t}(\gamma_a v_a^i)$ and $\frac{\delta \lambda}{\delta y_a^i} = \frac{\partial
\lambda}{\partial y_a^i}-\frac{\ud}{\ud t} ( \frac{\partial \lambda}{\partial
v_a^i} )$, and the dots refer to time derivatives. The potential $V$ (such that $E = \sum_a m_a c^2 \gamma_a + V$) reads 
\begin{equation}\label{V}
V = \sum_a v_a^i \frac{\partial \lambda}{\partial v_a^i} - \lambda - \sum_a v_a^i \dot{q}_a^i \,.
\end{equation}
The left-hand sides of~\eqref{eomexpl} and~\eqref{V} are known from Eqs.~\eqref{eom1}--\eqref{eom2} and~\eqref{potV}. However, the two equations are not independent, since $f_a^i$ and $V$ satisfy the constraint
\begin{equation}\label{constraint}
\frac{\ud V}{\ud t} + \sum_a v_a^i f_a^i = 0\,.
\end{equation}

In order to find $L$, our strategy is to determine first a particular Lagrangian $\hat{L}$, characterized by $(\hat{\lambda}, \hat{q}_a^i)$, such that the conjugate momenta $\hat{q}_a^i$ obey 
\begin{equation}\label{equation}
\sum_a v_a^i\hat{q}_a^i=0 \,.
\end{equation}
To order $G$, the same equation is also satisfied by the time derivative $\dot{\hat{q}}_a^i$. Therefore, for the particular solution $(\hat{\lambda}, \hat{q}_a^i)$, the equation~\eqref{V} reduces to an ordinary Legendre transformation,
\begin{equation}\label{PDE}
V = \sum_a v_a^i \frac{\partial \hat{\lambda}}{\partial v_a^i} - \hat{\lambda} \,.
\end{equation}
To determine $\hat{\lambda}$, we note that the potential $V$ given by~\eqref{conservedE} reduces in the limit $c\to +\infty$ to the Newtonian approximation, namely $V = U + \mathcal{O}(1/c^2)$ where
\begin{equation}\label{VN}
U = - \sum_{a<b} \frac{G m_a m_b}{r_{ab}} \,.
\end{equation}
If we subtract its Newtonian limit $U$ to $V$, we get a quantity which tends to zero when $c\to +\infty$ like $\mathcal{O}(1/c^2)$. Then, it is straightforward to show that a well-behaved solution of Eq.~\eqref{PDE} is
\begin{equation}\label{lambdasol}
\hat{\lambda} = - U + \frac{1}{c}\int_{c}^{+\infty} \ud s \biggl[ V\Bigl(\bm{y}_a, \frac{\bm{v}_a}{s}\Bigr) - U(\bm{y}_a)\biggr]\,.
\end{equation}
Namely, we have to insert into Eq.~\eqref{potV} all the relevant factors $c$ and make the replacement of $c$ by $s$, then integrate over the ``speed of light'' $s$ from the physical value $c$ up to infinity. The bound $s\to+\infty$ of the integral corresponds to the Newtonian limit and we see from the definition of the Newtonian potential~\eqref{VN} that the integral is convergent. The first term in Eq.~\eqref{lambdasol} represents the Newtonian approximation with the correct minus sign for a Lagrangian, and the integral represents formally the complete series of PN corrections, but resummed in the PM approximation. The result~\eqref{lambdasol} can be rewritten in a simpler way as the Hadamard ``partie finie'' (Pf) of the integral, in the same sense as was used in Eq.~\eqref{Stail}, for taking care of the divergence at infinity:
\begin{equation}\label{lambdasolHad}
\hat{\lambda} = \text{Pf} \,\frac{1}{c}\int_{c}^{+\infty} \ud s \,V\Bigl(\bm{y}_a, \frac{\bm{v}_a}{s}\Bigr) \,.
\end{equation}
For this very simple type of divergence $\sim s^0 + \mathcal{O}(s^{-2})$ the Pf does not depend on an arbitrary constant, unlike in~\eqref{Stail}. The expressions~\eqref{lambdasol}--\eqref{lambdasolHad} give a particular solution of the equation~\eqref{V} but we still have to adjust $\hat{q}_a^i$ in order to satisfy the EOM, see~\eqref{eomexpl}. Thus, we look for $\hat{q}_a^i$ satisfying 
\begin{equation}\label{eomexpl2}
\ddot{\hat{q}}_a^i = f_a^i - \frac{\delta \hat{\lambda}}{\delta y_a^i} \,,
\end{equation}
where the right-hand side is known. To order $G$ we have been able to integrate twice this relation to determine $\hat{q}_a^i$, and that solution automatically satisfies the constraint~\eqref{equation} by virtue of~\eqref{constraint}.

Finally we have found a particular Lagrangian $(\hat{\lambda}, \hat{q}_a^i)$. Now the general solution $(\lambda, q_a^i)$ can be obtained by adding an arbitrary total time-derivative $\ud F/\ud t$, where $F$ is a function of the positions $y_a^i$ and velocities $v_a^i$. Hence the general solution (for the class of harmonic-coordinate Lagrangians that are linear in accelerations) reads
\begin{subequations}\label{gaugefreedom}
\begin{align}
\lambda &= \hat{\lambda} + \sum_a v_a^i \frac{\partial F}{\partial y_a^i} \,,\\
q_a^i &= \hat{q}_a^i + \frac{\partial F}{\partial v_a^i} \,.
\end{align}\end{subequations}
At 1PM order the Lagrangian in harmonic coordinates irreducibly depends on accelerations, \textit{i.e.}, it is impossible to determine $F$ such that $q_a^i=0$. However, we know that the accelerations in a Lagrangian can be eliminated by appropriate shifts of the trajectories. In fact, it can be shown that the particular solution $\hat{\lambda}$ found in~\eqref{lambdasol}--\eqref{lambdasolHad} represents an ordinary Lagrangian which is physically equivalent but expressed in some shifted (non harmonic) variables.\cite{BFok18} 

Given the complicated structure of $V$ in Eq.~\eqref{potV}, we could not find a closed form expression for the 1PM harmonic coordinate Lagrangian in the general case. However, we could easily work out the integral~\eqref{lambdasol} in the PN approximation $c\to+\infty$ to any order. We start from the known 4PN expansion of the potential $V$ following from~\eqref{potV}, and explicitly perform the integration~\eqref{lambdasol} term by term, to obtain the corresponding 4PN expansion of $\hat{\lambda}$. Then, we derive the coefficient of accelerations $\hat{q}_a^i$ at 4PN order, see\cite{BFok18} for details. Finally, we find a unique total time-derivative, with some function $F_\text{PN}$ given in the form of a PN expansion, so that the Lagrangian satisfyingly agrees up to order $G$ with the published 4PN Lagrangian.\cite{BBBFMc,BBFM17} Furthermore, we have pushed the analysis to the next order and obtained all the terms of order $G$ in the harmonic coordinates Lagrangian up to the 5PN order.\cite{BFok18} 

In another application, we worked out the case of equal masses for which it is possible to find a closed form expression for the Lagrangian, and we have verified that the associated Hamiltonian differs from the one obtained by Ledvinka, Sch{\"a}fer and Bi\v{c}\'{a}k~\cite{LSB08} in the ADM Hamiltonian by a mere canonical transformation.

\bibliography{ListeRef_MG15.bib}

\end{document}